\documentclass[10pt,letterpaper]{article}
\usepackage{amssymb}
\usepackage{opex3}
\usepackage{subfigure}
\usepackage{graphicx}

\begin{document}

\title{Gap solitons in grating superstructures}
\author{Thawatchai Mayteevarunyoo,$^{1*}$ and Boris A. Malomed$^{2}$}
\address{$^{1}$Department of Telecommunication Engineering,
Mahanakorn University of Technology, \\Bangkok 10530, Thailand\\
$^{2}$Department of Physical Electronics, School of Electrical
Engineering, \\Faculty of Engineering, Tel Aviv University, Tel Aviv 69978,
Israel\\
$^{*}$Corresponding author:
\underline{thawatch@mut.ac.th}}
\begin{abstract}
We report results of the investigation of gap solitons (GSs) in the generic
model of a periodically modulated Bragg grating (BG), which includes
periodic modulation of the BG chirp or local refractive index, and periodic
variation of the local reflectivity. We demonstrate that, while the
previously studied reflectivity modulation strongly destabilizes all
solitons, the periodic chirp modulation, which is a novel feature,
stabilizes a new family of double-peak fundamental BGs in the side bandgap
at negative frequencies (gap No. $-1$), and keeps solitons stable in the
central bandgap (No. $0$). The two soliton families demonstrate bistability,
coexisting at equal values of energy. In addition, stable 4-peak bound
states are formed by pairs of fundamental GSs in bandgap $-1$. Self-trapping
and mobility of the solitons are studied too.
\end{abstract}







\section{Introduction and the model}

The technology for writing grating superstructure (alias superlattices) on
optical fibers had become available twenty years ago \cite{Russell,Eggleton}%
. These superlattices are Bragg gratings (BGs) with a long-wave modulation
of period $\sim 1$ mm imposed on them, while the underlying BG period is $%
\lambda /2\lesssim 1$ $\mu $m ($\lambda $ is the wavelength of light coupled
into the BG). A theoretical model shows that, in addition to the central
bandgap generated by the underlying uniform BG, the superstructure gives
rise to a new set of bandgaps \cite{theory}. In this connection, it is
relevant to mention that the modulation of the periodic lattice potential in
the Schr\"{o}dinger equation, produced by beatings between two lattices with
close periods, also gives rise to additional narrow ``mini-gaps" in the
respective spectrum \cite{minigap}. Taking into regard the Kerr nonlinearity
of the fiber, as in the theory of gap solitons (GSs) in the uniform fiber BG
\cite{SterkeSipe,Aceves,Christo}, ``coupled-supermode" equations were
derived in Ref. \cite{theory}, and examples of the corresponding GSs were
found (these equations bear a similarity to coupled-mode equations for deep
BGs \cite{deep1,deep2}). Stable solitons in the above-mentioned mini-gaps of
the Gross-Pitaevskii equation with the repulsive cubic nonlinearity, which
is a model of the Bose-Einstein condensate (BEC) trapped in the optical
lattice, were found too \cite{minigap}. Another example of the
superstructure was developed in the form of the Moir\'{e} pattern, with a
sinusoidal modulation imposed on the periodic variation of the refractive
index underlying the ordinary BG. The Moir\'{e} supergrating features a
narrow transmission band in the middle of the central gap, which was
proposed \cite{Khurgin,Moire2} and realized experimentally \cite{SlowMoire}
as a means for the retardation of light in gratings.

Cellular optical media which resemble the BG structure and may also be used
as a basis for the design of superstructures are CROWs (coupled resonant
optical waveguides) \cite{CROWtoo,CROW}. It is also possible to realize
superstructure patterns in the recently proposed ``semi-discrete" BG (a
waveguide with uniform nonlinearity and periodically distributed short
segments with strong Bragg reflectivity) \cite{Kobi}. A vast potential for
the synthesis of complex grating patterns is offered by techniques developed
for writing BGs in photonic crystals and photonic-crystal fibers \cite%
{PhotCryst,PhotCryst2,PhotCryst3}.

A topic of fundamental significance is families of GSs and their stability
in models describing superstructured BGs. In fact, the stability of GSs is a
nontrivial issue even in the standard model of uniform BGs \cite%
{Rich,CapeTown,Trillo}). A basic system of coupled-mode equations for
counterpropagating waves $u(x,t)$ and $v(x,t)$ in the periodically modulated
BG was proposed in Ref. \cite{we}. In the normalized form, it is%
\begin{eqnarray}
i\frac{\partial u}{\partial t}+i\frac{\partial u}{\partial x}+\left[
1-\varepsilon \cos \left( kx\right) \right] v+\mu \cos \left( kx\right)
\cdot u+\left( \left\vert v\right\vert ^{2}+\frac{1}{2}\left\vert
u\right\vert ^{2}\right) u &=&0,  \nonumber \\
&&  \label{CME} \\
i\frac{\partial v}{\partial t}-i\frac{\partial v}{\partial x}+\left[
1-\varepsilon \cos \left( kx\right) \right] u+\mu \cos \left( kx\right)
\cdot v+\left( \left\vert u\right\vert ^{2}+\frac{1}{2}\left\vert
v\right\vert ^{2}\right) v &=&0.  \nonumber
\end{eqnarray}%
Here, $\varepsilon $ is the amplitude of the modulation of the Bragg
reflectivity (in other words, it represents periodic \textit{apodization} of
the grating \cite{Tsoy1}), while $\mu $ admits two interpretations: it
accounts for the periodic variation of the local \textit{chirp} of the BG
\cite{Tsoy2,we}, or of the effective refractive index of the carrying fiber.
The spatial period of both modulations is $2\pi /k$. We define the model by
fixing $\mu >0$, while $\varepsilon $ may take zero, positive, and negative
values.

It is known that GSs are possible not only as temporal solitons in fiber
gratings, but also as spatial solitons in planar waveguides equipped with
the grating in the form of a system of parallel grooves \cite%
{spatial1,spatial2,spatial3,KA}, as well as solitons in photonic crystals
\cite{spatial4}. Equations (\ref{CME}) may also be interpreted in that
context (replacing $t$ by propagation coordinate $z$), with $\mu $
representing the amplitude of a long-wave longitudinal modulation of the
refractive index in a layered planar waveguide.

The results obtained in this work are presented in Section II, where
families of soliton solutions and their stability are reported, and in
Section III, which deals with the self-trapping and nonlinear evolution of
stable and unstable GSs, and with moving solitons. In the previously studied
model of the reflectivity modulation \cite{we}, the GSs quickly become
unstable with the increase of modulation amplitude $\varepsilon $. In
Section II we demonstrate that the effect of the periodic modulation of the
local chirp (or refractive index) -- a feature that was not studied before
-- is different: a part of the GS family filling out the central bandgap
(labeled as gap $0$ below, see Fig. \ref{fig1}) remains stable with the
increase of $\mu $, while the first side bandgap emerging at $\omega <0$
(designated below as gap $-1$) supports a new partially stable family of
fundamental GSs, whose characteristic feature is a \emph{two-peak }shape,
unlike the ordinary single-peak solitons existing in the central bandgap (in
bandgap $+1$, GSs also feature the double-peak shape, but they are
unstable). Note that fundamental GSs in the Gross-Pitaevskii equation with a
periodic potential do not feature a dual-peak structure. In terms of the
spatial-domain model, the double-peak solitons may find an application as
optically induced conduits routing weak signal beams \cite{KA}. In Section
III it is shown that, in the model with $\varepsilon =0$ and $\mu >0$,
stable quiescent solitons belonging to the central gap readily self-trap
from \textit{moving} input pulses of a general form, hence the periodically
chirp-modulated BG may serve as a tool for the creation of solitons of
standing-light.

Unlike the standard BG model, in the present system stable double- and
single-peak solitons, residing in gaps $-1$ and $0$, respectively, feature
\emph{bistability}, coexisting at equal values of energy. 4-peak bound
states of two double-peak solitons, and 3-peak complexes, built of three
single-peak solitons, may be stable too (recall that bound states of GSs do
not exist in the standard BG).

In Section III we demonstrate that the evolution of unstable GSs in the
modulated system features another novelty: unstable solitons with a
sufficiently large energy self-retrap into stable double-peak GSs belonging
to bandgap $-1$, while unstable GSs do not transform themselves into stable
ones in the standard model. Other unstable GSs evolve into persistent
breathers, or may be destroyed by the instability. In Section III we also
study a possibility to set quiescent GSs in motion, which is suggested by
the fact that, thus far, BG solitons in fiber gratings have been created
only at finite velocity $c$; in the first works, it was $c\geq 0.5$ [with
respect to the largest velocity in Eqs. (\ref{CME}), $c_{\max }=1$] \cite%
{experiment1,experiment2}, while later it was brought down to $c\approx 0.23$
\cite{slow}. In terms of the above-mentioned spatial-domain interpretation,
moving solitons correspond to tilted beams. We demonstrate that stable
moving solitons are supported by Eqs. (\ref{CME}) with $\varepsilon =0$ and
small values of $\mu $. In fact, these results also stress that the
modulated BG offers a possibility to bring moving pulses to a halt and thus
create solitons of standing light.

\section{Stationary solutions and their stability}

\subsection{The mode of the analysis}

Stationary solutions of Eqs. (\ref{CME}) with frequency $\omega $ and zero
velocity are looked for as $\left\{ u\left( x,t\right) ,v\left( x,t\right)
\right\} =\left\{ U\left( x\right) ,V\left( x\right) \right\} \exp \left(
-i\omega t\right) $, with complex functions $U$ and $V$ satisfying equations%
\begin{eqnarray}
&&+i\frac{dU}{dx}+\left[ \omega +\mu \cos \left( kx\right) \right] U+\left[
1-\varepsilon \cos \left( kx\right) \right] V+\left[ \left( \left\vert
V\right\vert ^{2}+\frac{1}{2}\left\vert U\right\vert ^{2}\right) \right] U=0,
\nonumber \\
&&  \label{Solutions} \\
&&-i\frac{dV}{dx}+\left[ \omega +\mu \cos \left( kx\right) \right] V+\left[
1-\varepsilon \cos \left( kx\right) \right] U+\left[ \left( \left\vert
U\right\vert ^{2}+\frac{1}{2}\left\vert V\right\vert ^{2}\right) \right] V=0.
\nonumber
\end{eqnarray}%
For the numerical solution, the complex amplitudes were split into real and
imaginary parts, $\left\{ U(x),V(x)\right\} \equiv \left\{
U_{1}(x),V_{1}(x)\right\} +i\left\{ U_{2}(x),V_{2}(x)\right\} $, and the
resulting system of four equations was solved by means of the Newton's
iteration method. The initial guess generating even solutions was $%
U_{10}\left( x\right) =U_{20}\left( x\right) =V_{10}\left( x\right)
=V_{20}\left( x\right) =A~\mathrm{sech}\left( ax\right) $, with constants $A$
and $a$.

Numerical results are reported below for $k=1$, which represents the generic
situation. Families of soliton solutions are characterized by the total
energy (on total power, in terms of the spatial-domain model),
\begin{equation}
E=\int_{-\infty }^{+\infty }\left( \left\vert u\right\vert ^{2}+\left\vert
v\right\vert ^{2}\right) dx,  \label{E}
\end{equation}%
to be presented as a function of $\omega $.

\begin{figure}[h]
\centering\subfigure[]{\includegraphics[width=2.5in]{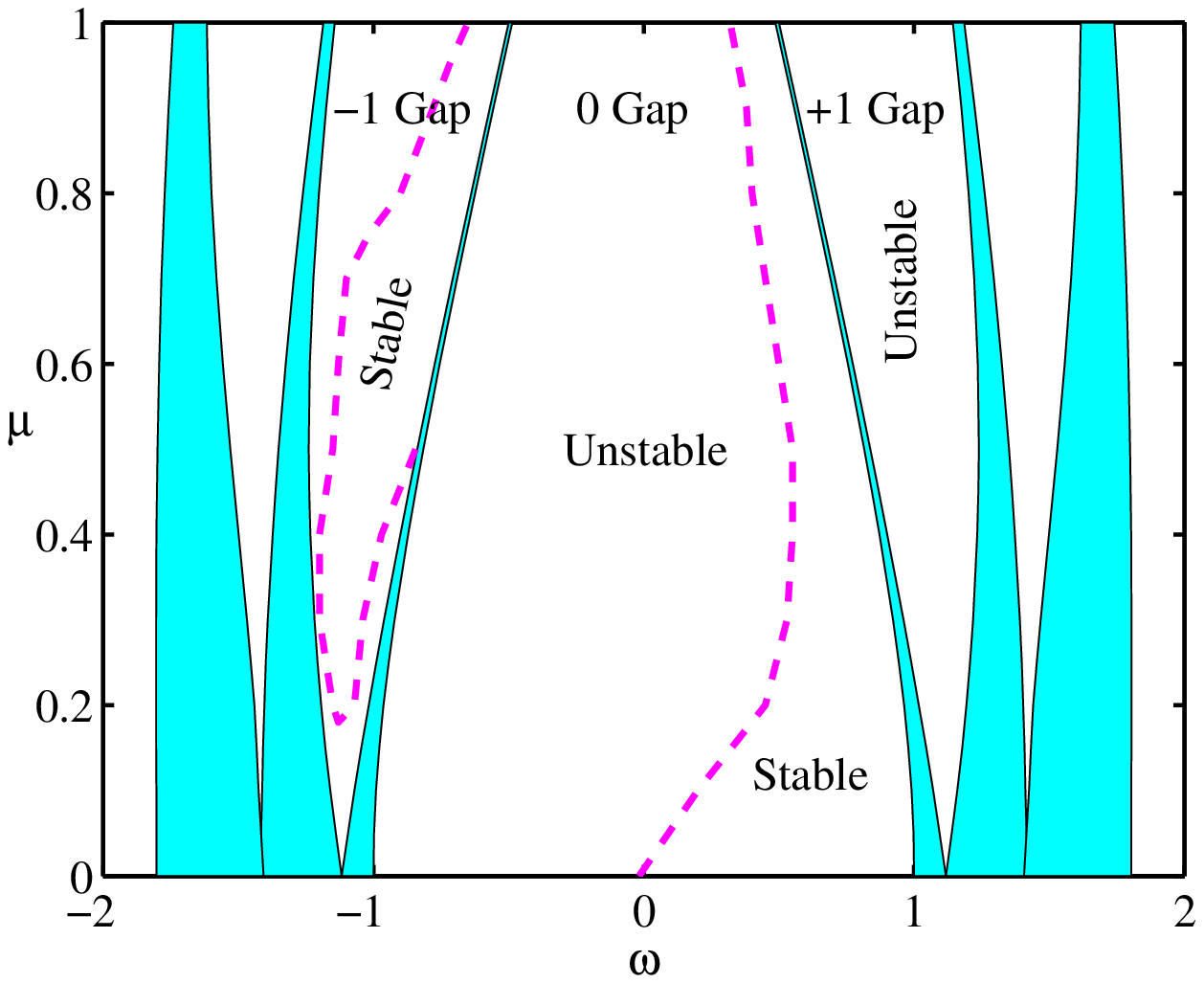}}%
\subfigure[]{\includegraphics[width=2.5in]{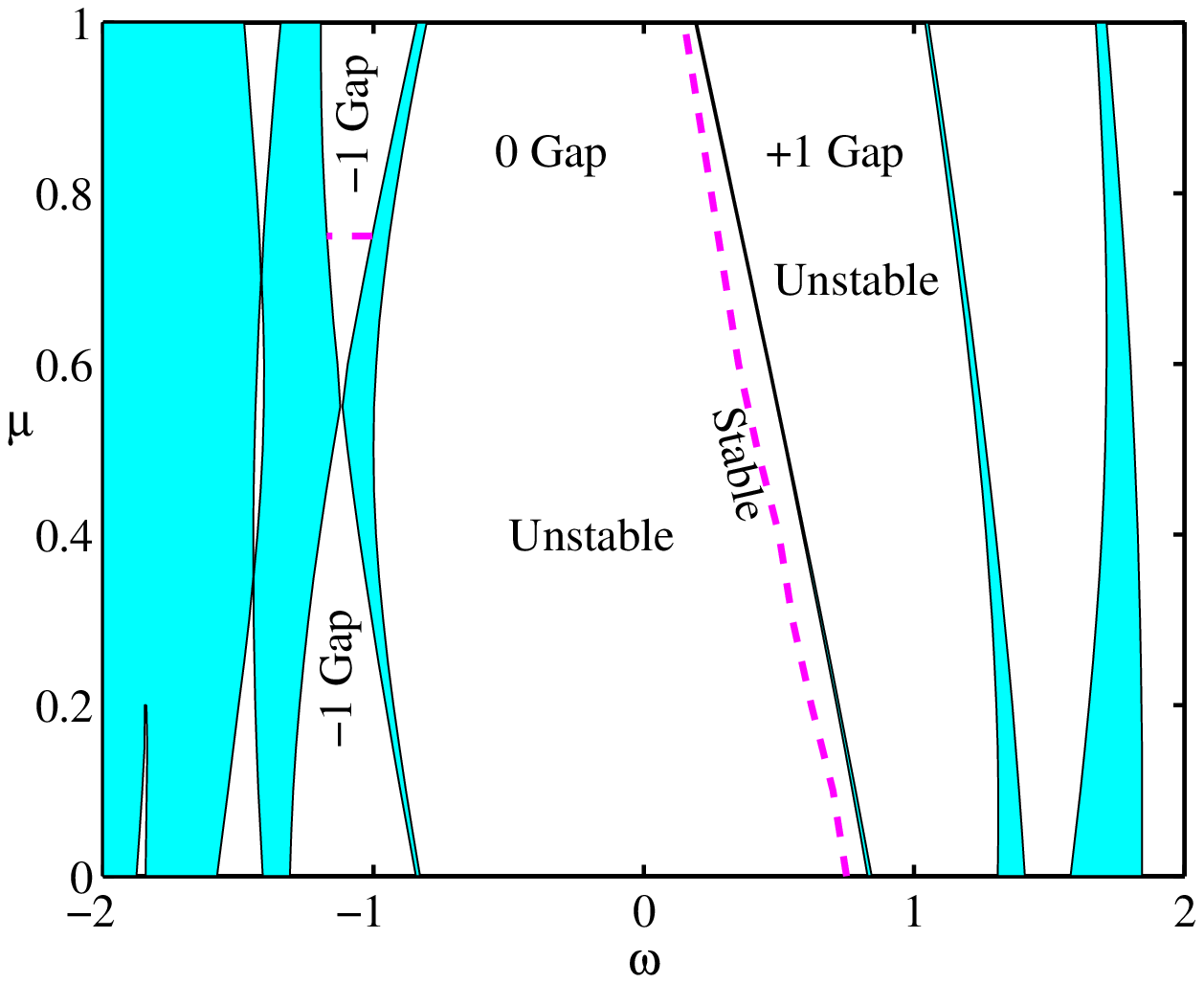}} \subfigure[]{%
\includegraphics[width=2.5in]{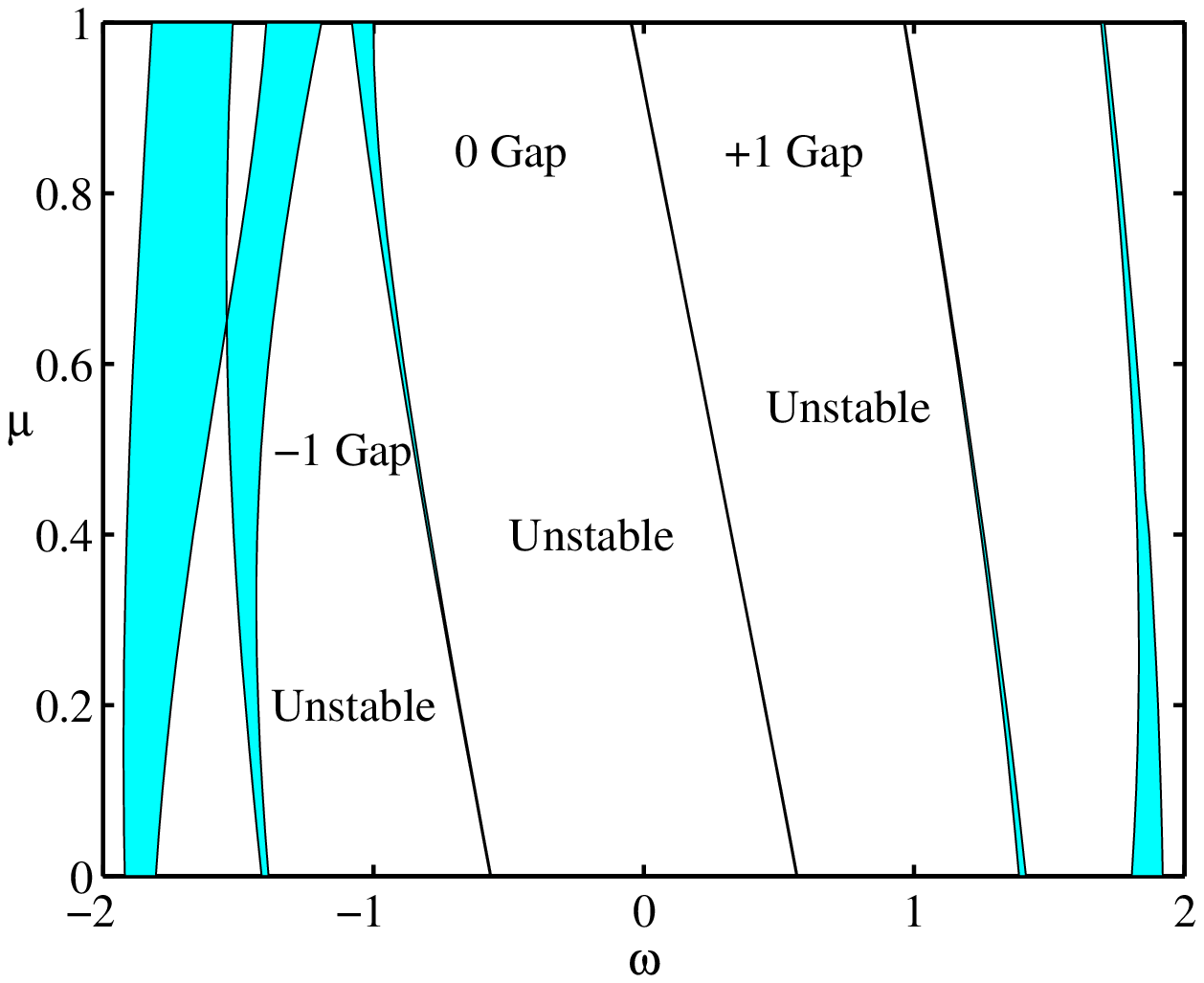}}\subfigure[]{%
\includegraphics[width=2.5in]{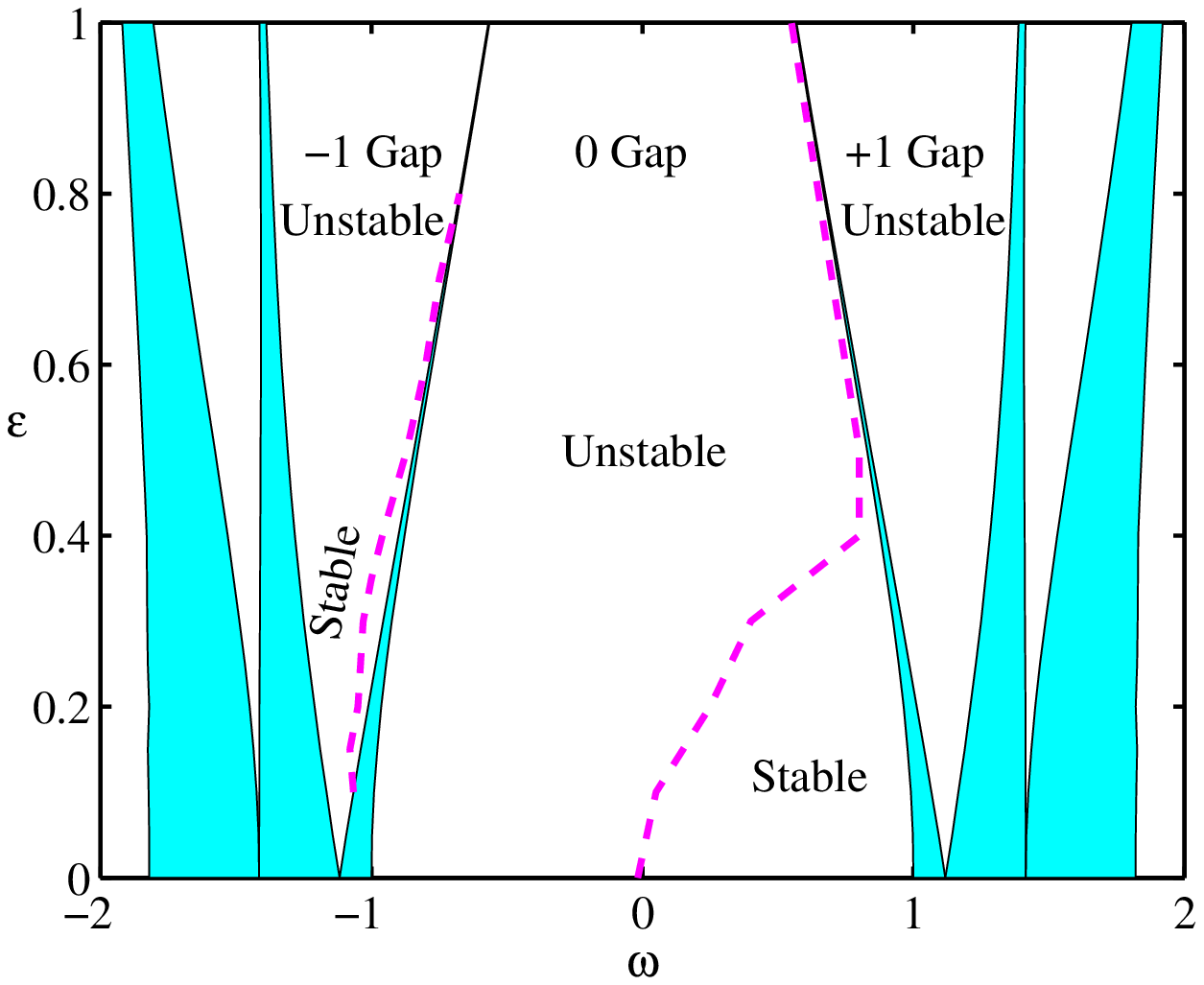}}
\caption{The bandgap structure found from the linearization of Eqs. (\protect
\ref{Solutions}) for (a) $\protect\varepsilon =0$, (b) $\protect\varepsilon %
=0.5$, (c) $\protect\varepsilon =1$ and (d) $\protect\mu =0$. Shaded areas
are occupied by Bloch bands. Five gaps are displayed: the central one (No. $%
0 $) and two side bandgaps, $\pm 1$ and $\pm 2$ (gaps $\pm 2$ are not
labeled). Stable solitons are found in gaps $0$ and $-1$, where borders
between stability and instability areas are shown by dashed lines. Note that
all solitons are unstable for $\protect\varepsilon =1$.}
\label{fig1}
\end{figure}

The bandgap spectrum of the linearized version of Eqs. (\ref{Solutions}) was
computed by means of software package \texttt{SpectrUW} \cite{SpectrUW}. The
spectra are displayed in Fig. \ref{fig1}, which also show stability borders
of GS families found in the bandgaps from the solution of the full nonlinear
system, as described below. Note that the region occupied by bandgap $-1$ in
Fig. \ref{fig1}(b) (for $\varepsilon =0.5$) splits into two parts, with
stable solitons existing only in the upper one.

The linearization of Eqs. (\ref{Solutions}) is invariant with respect to
transformation $\varepsilon \rightarrow -\varepsilon $, $\omega \rightarrow
-\omega $, $x\rightarrow x+\pi /k$, $\left\{ U,V\right\} \rightarrow \left\{
-V,U\right\} $; therefore, the linear spectrum for $\varepsilon <0$ can be
obtained as a mirror image (with $\omega \rightarrow -\omega $) from its
counterpart for $-\varepsilon $. However, this transformation does not apply
to full nonlinear equations (\ref{Solutions}). On the other hand, Eqs. (\ref%
{Solutions}) admit the reduction to a single equation by means of the
well-known substitution, $V=\pm U^{\ast }$. As well as in the standard
model, the GSs found in the central bandgap satisfy ``ordinary" reduction $%
V=-U^{\ast }$, while double-peak solitons populating bandgap $-1$ (and
unstable solitons of the same type in bandgap $+1$) obey the
``extraordinary" reduction, $V=U^{\ast }$.

Stability of solitons was identified by dint of simulations of the evolution
of perturbed solitons, typically up to $t=10,000$, which means several
thousand soliton periods, or time $\sim 1$ ns, in physical units. It was
additionally checked, in typical cases, that the solutions which are
identified as stable ones retain their stability\ in arbitrarily long
simulations. The simulations were performed by means of the split-step
Fourier-transform method, with absorbers placed at edges of the integration
domain. The domain was covered by a mesh consisting of $N=\allowbreak 512$
grid points, and the stepsize of the time integration was $\Delta t=0.01$
(it was checked that further increase of $N$ and decrease of $\Delta t$ did
not alter the results).

Figure 1 clearly shows that the increase of the reflectively modulation,
represented by $\varepsilon $, quickly destabilizes all solitons. On the
other hand, the model with the periodic chirp modulation, which is accounted
for by $\mu $ (unlike the system with $\varepsilon >0$, it was not studied
before), supports stable GSs, including the new family in gap $-1$.
Therefore, we focus below on the study of this model; some new results for
the case of $\mu =0$ and $\varepsilon \neq 0$ will be included too, for the
sake of comparison.

\subsection{Results}

\begin{figure}[h]
\centering\subfigure[]{\includegraphics[width=2.5in]{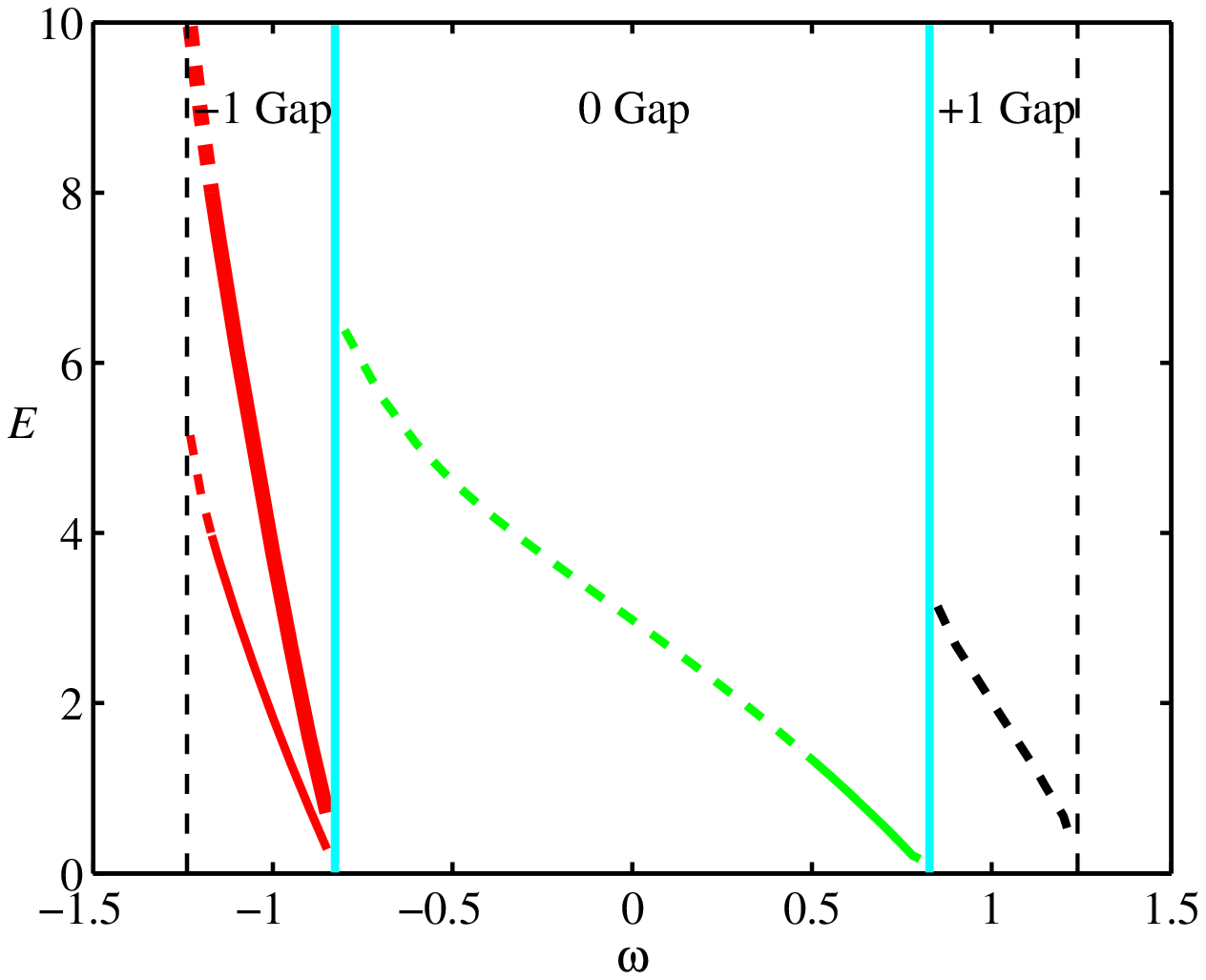}}%
\subfigure[]{\includegraphics[width=2.5in]{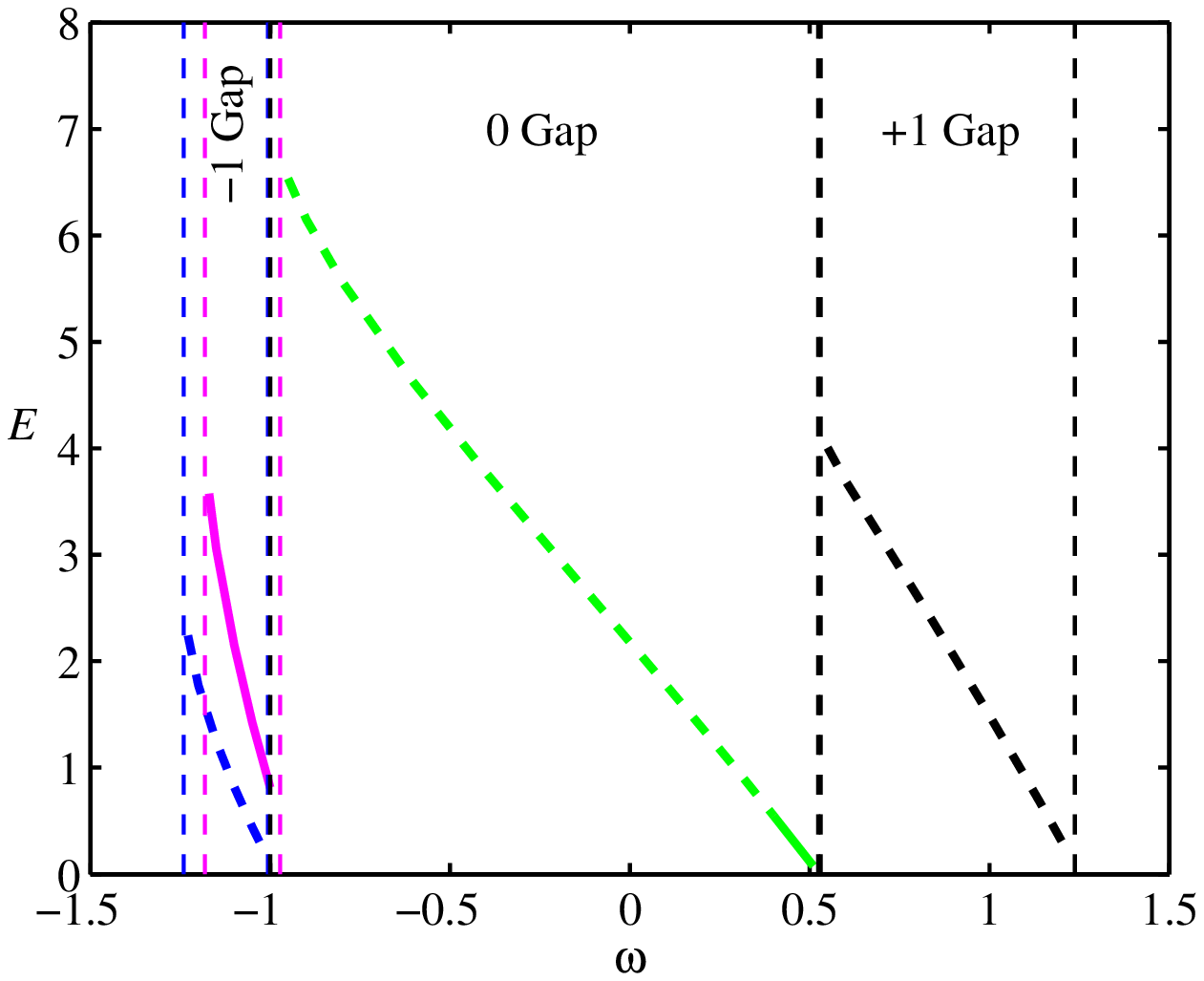}} \subfigure[]{%
\includegraphics[width=2.5in]{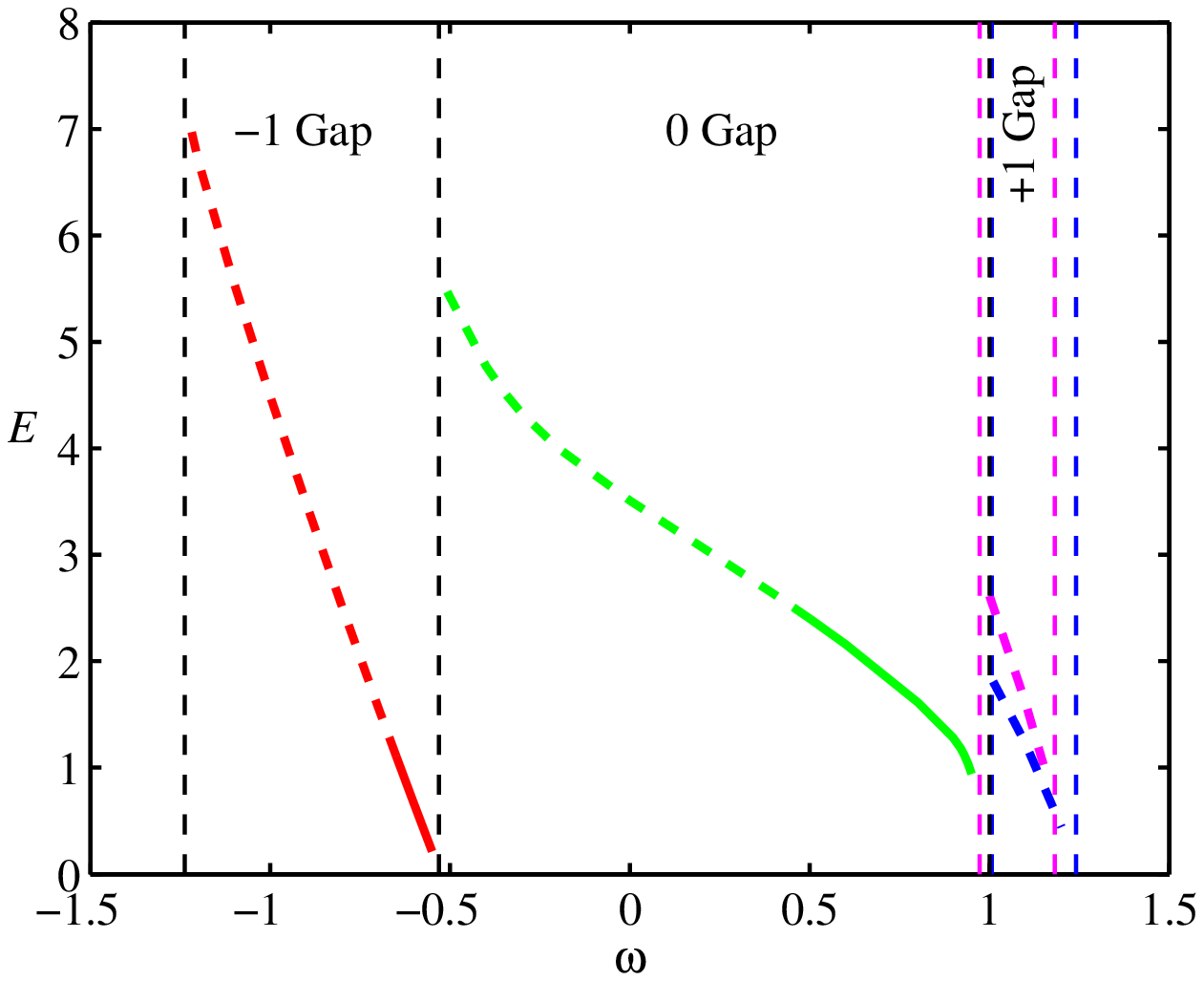}}
\caption{Gap-soliton families, shown in the form of energy $E$ versus
intrinsic frequency $\protect\omega $, for (a) $\protect\mu =0.5$, $\protect%
\varepsilon =0$, (b) $\protect\mu =\protect\varepsilon =0.5$, (c) $\protect%
\mu =-\protect\varepsilon =0.5$. Stable and unstable portions of the
families are depicted by continuous and dashed lines, respectively. The
upper bold curve in gap $-1$ in (a) represents the family of 4-peak bound
states of fundamental solitons. Two different curves in (b) and (c), in gaps
$-1$ and $+1$, respectively, pertain to two regions in which these gaps
exist, cf. Fig. \protect\ref{fig1}(b). Recall that, for $\protect\varepsilon %
<0$ [as in (c)], the bandgap structure is obtained from that for $-\protect%
\varepsilon $ as the mirror image, with $\protect\omega \rightarrow -\protect%
\omega $.}
\label{fig2}
\end{figure}

In addition to Fig. \ref{fig1}, the stability of the GSs is summarized in
Fig. \ref{fig2}, which displays typical dependences $E(\omega )$ [recall $E$
is defined in Eq. (\ref{E})] for soliton families in several generic cases
and in different bandgaps (situations where all solitons are unstable, such
as at $\varepsilon =1$, are not included). As said above, stable solitons
are found only in bandgaps $0$ and $-1$. For instance, the stability
intervals in gaps $-1$ and $0$ for $\mu =0.5$ and $\varepsilon =0$ are $%
-1.17<\omega <-0.84$ and $0.55<\omega <0.82$, respectively. If the existence
range of gap $-1$ splits into two parts, as in Fig. \ref{fig1}(b), stable
solitons are found only in the upper one [in Fig. \ref{fig1}(b), the
stability area in bandgap $-1$ is located at $\mu >0.75$)]. A notable
feature observed in Figs. \ref{fig2}(a,c) is the \textit{bistability}:
stable portions of the GS families in gaps $0$ and $-1$ may cover identical
intervals of energy. In higher-order bandgaps, starting from $\pm 2$, all
GSs are unstable.

\begin{figure}[h]
\centering\subfigure[]{\includegraphics[width=2.5in]{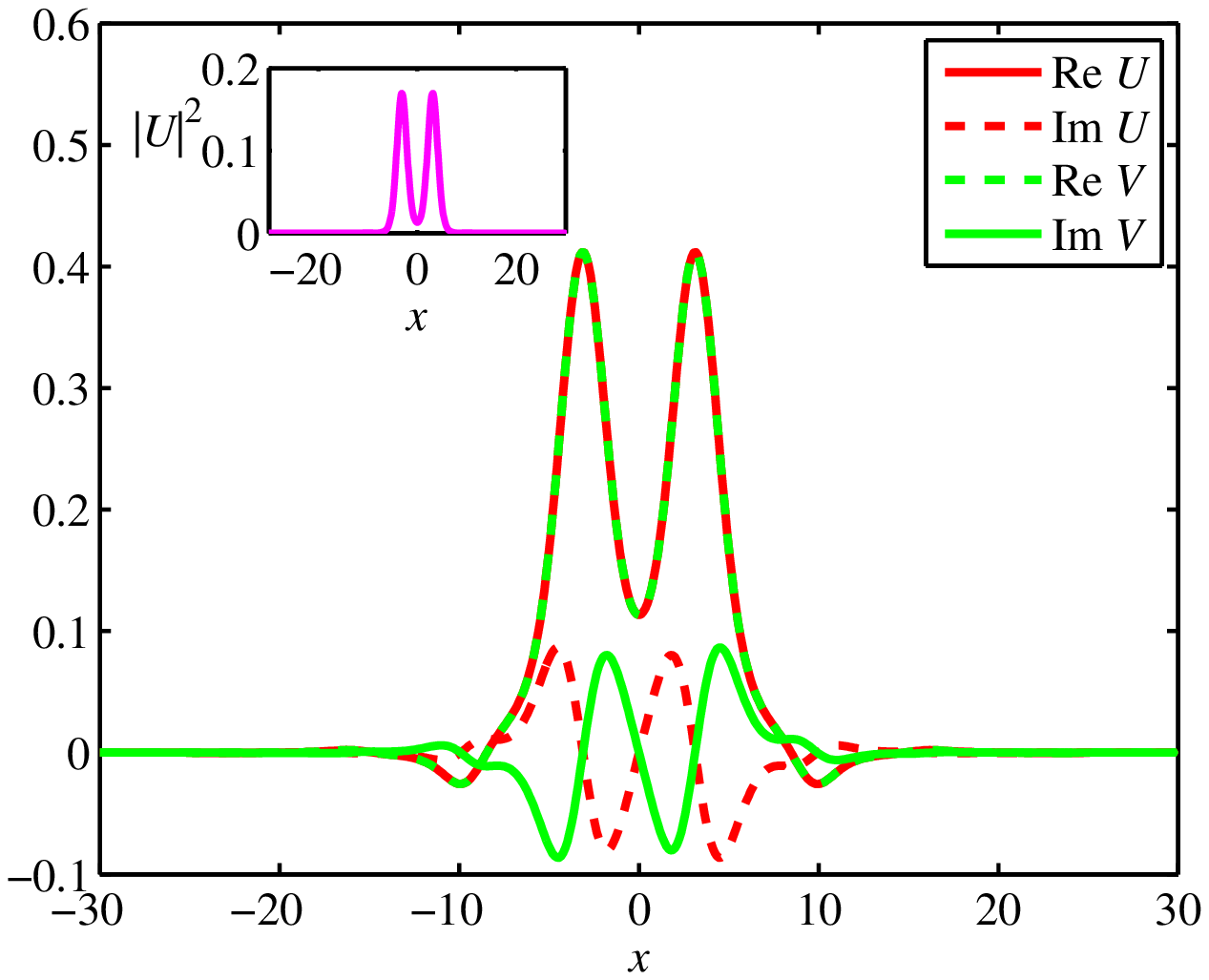}}%
\subfigure[]{\includegraphics[width=2.5in]{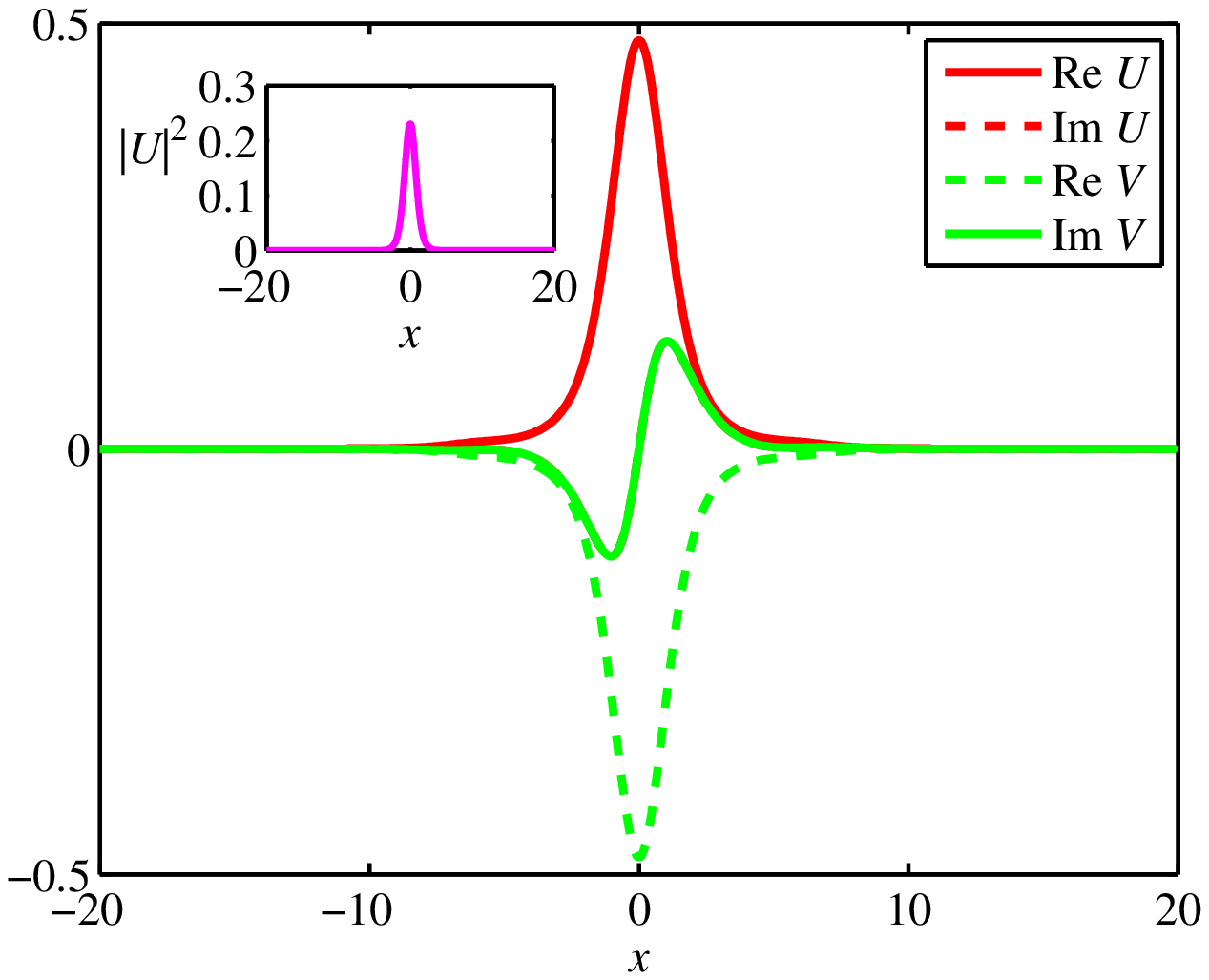}}
\caption{(a) A stable double-peak soliton found in gap $-1$, for $\protect%
\mu =0.5$, $\protect\varepsilon =0$, and $\protect\omega =-1$. The energy of
this soliton is $E=1.83$. (b) A stable single-peak soliton in gap $0$, for $%
\protect\mu =0.5$, $\protect\varepsilon =0$, and $\protect\omega =0.6$. Its
energy is $E=0.96$.}
\label{fig3}
\end{figure}

A characteristic feature of the GSs in bandgap $-1$ is the double-peak
shape, as shown in Fig. \ref{fig3}(a). We stress that the double-peak GSs
are fundamental solitons, rather than bound states of some single-peak
pulses. Note that all GSs in bandgap $-1$ have a single peak in the model
with $\mu =0$ and $\varepsilon >0$ \cite{we} [and almost all of them are
unstable, see Fig. \ref{fig1}(d)]. As mentioned above, the solitons in gap $%
-1$obey the ``extraordinary" reduction, $V=U^{\ast }$. Unlike them, in gap $0
$ GSs are similar to their counterparts in the standard model, being subject
to the ordinary reduction, $V=-U^{\ast }$, see Fig. \ref{fig3}(b).

\begin{figure}[h]
\centering\subfigure[]{\includegraphics[width=2.5in]{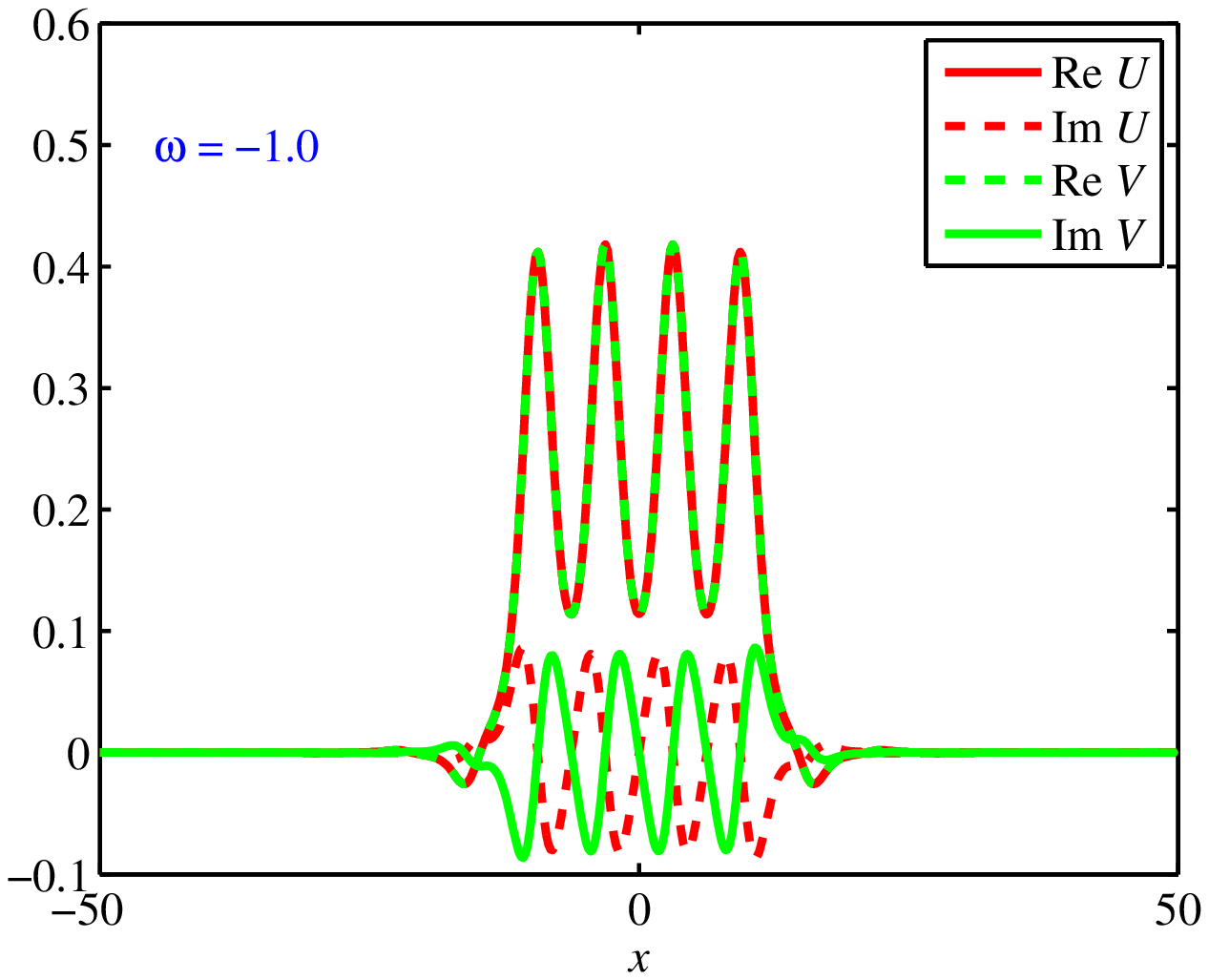}}%
\subfigure[]{\includegraphics[width=2.5in]{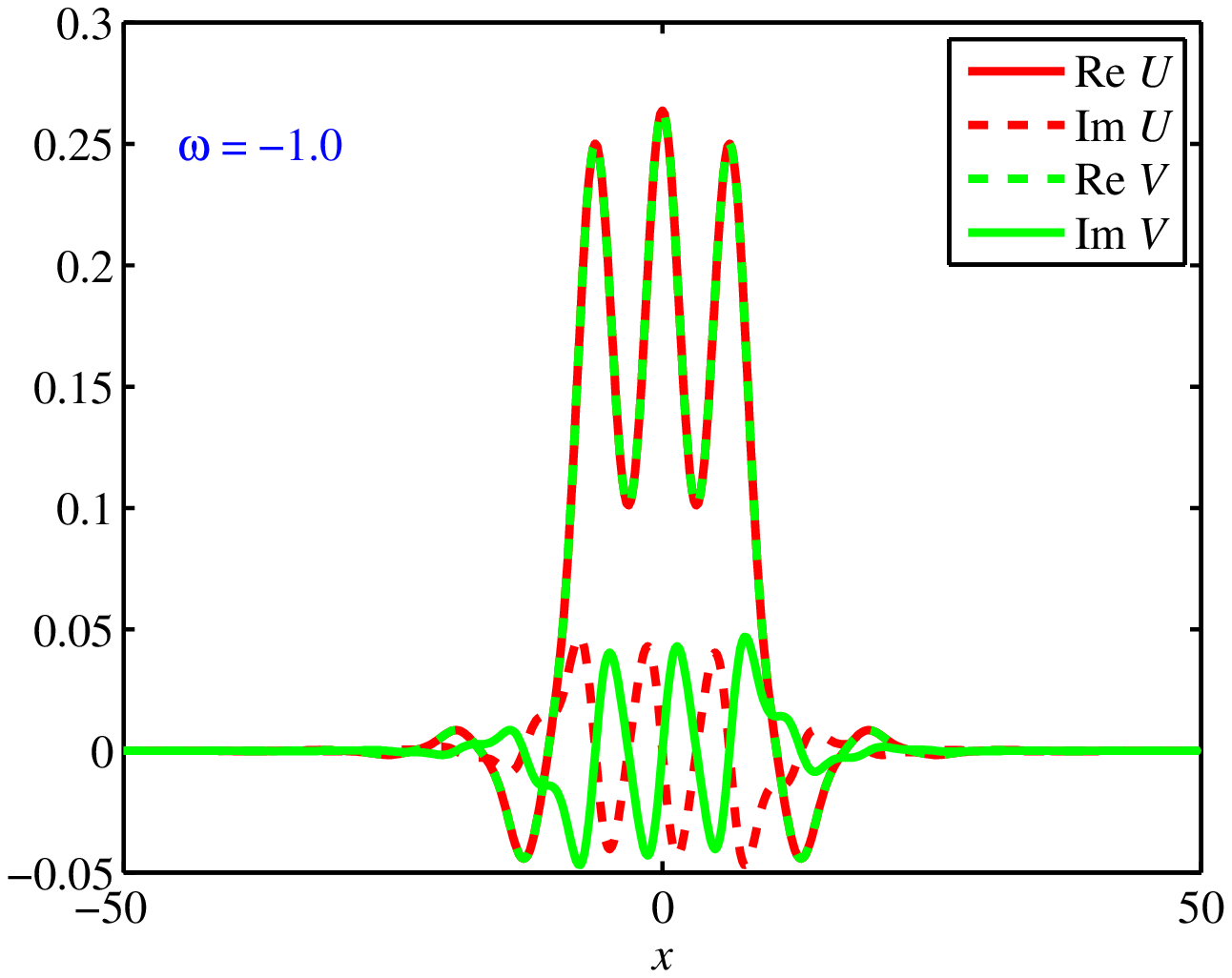}}
\caption{(a) A stable bound state of two fundamental twin-peak solitons in
gap $-1$, for $\protect\mu =0.5$, $\protect\varepsilon =0$, and $\protect%
\omega =-1.0$. The energy of this state is $3.76$, while the energy of each
constituent soliton is $1.83$. (b) A stable bound state of three single-peak
solitons for $\protect\mu =0$, $\protect\varepsilon =0.3$, and $\protect%
\omega =-1$. The energy of the bound state is $1.30$, the energy of each
constituent being $0.35$.}
\label{fig4}
\end{figure}

Unlike the GSs in the central bandgap, which do not combine into bound
states, solitons in bandgap $-1$ may form several species of complexes,
symmetric and anti-symmetric ones. Only one of them is stable, \textit{viz}%
., a 4-peak symmetric bound state of two double-peak solitons, see an
example in Fig. \ref{fig4}(a). The entire family of such states is shown in
Fig. \ref{fig2}(a) by the upper bold curve. The conclusion that the 4-peak
states are bound states of fundamental solitons is clearly suggested by the
comparison of curves $E(\omega ),$ which shows that the energy of the 4-peak
structure is, approximately, twice that of the double-peak soliton at the
same $\omega $. The stability area of the 4-peak states is identical to that
of the fundamental GSs. In addition, stable 3-peak symmetric bound states of
three single-peak solitons were found in that small part of gap $-1$ in the
model with $\varepsilon \neq 0$ and $\mu =0$ where the GSs are stable as per
Fig. \ref{fig1}(d), see an example in Fig. \ref{fig4}(b) (bound states were
not studied in Ref. \cite{we}).

\section{Nonlinear evolution of stable and unstable solitons}

\subsection{Self-trapping of stable solitons}

To appraise the experimental relevance of the GSs, it is necessary to
consider the possibility of self-trapping of such solitons from standard
input pulses (Gaussians). In the fiber BG, the input always has a finite
velocity, and, obviously, it may contain only a single (forward) component.
In the spatial-domain setting, the input beams may be both straight and
tilted (the former one corresponds to zero velocity in the temporal domain),
and simultaneous coupling of both components into the grating is possible.

\begin{figure}[h]
\centering\subfigure[]{\includegraphics[width=2.5in]{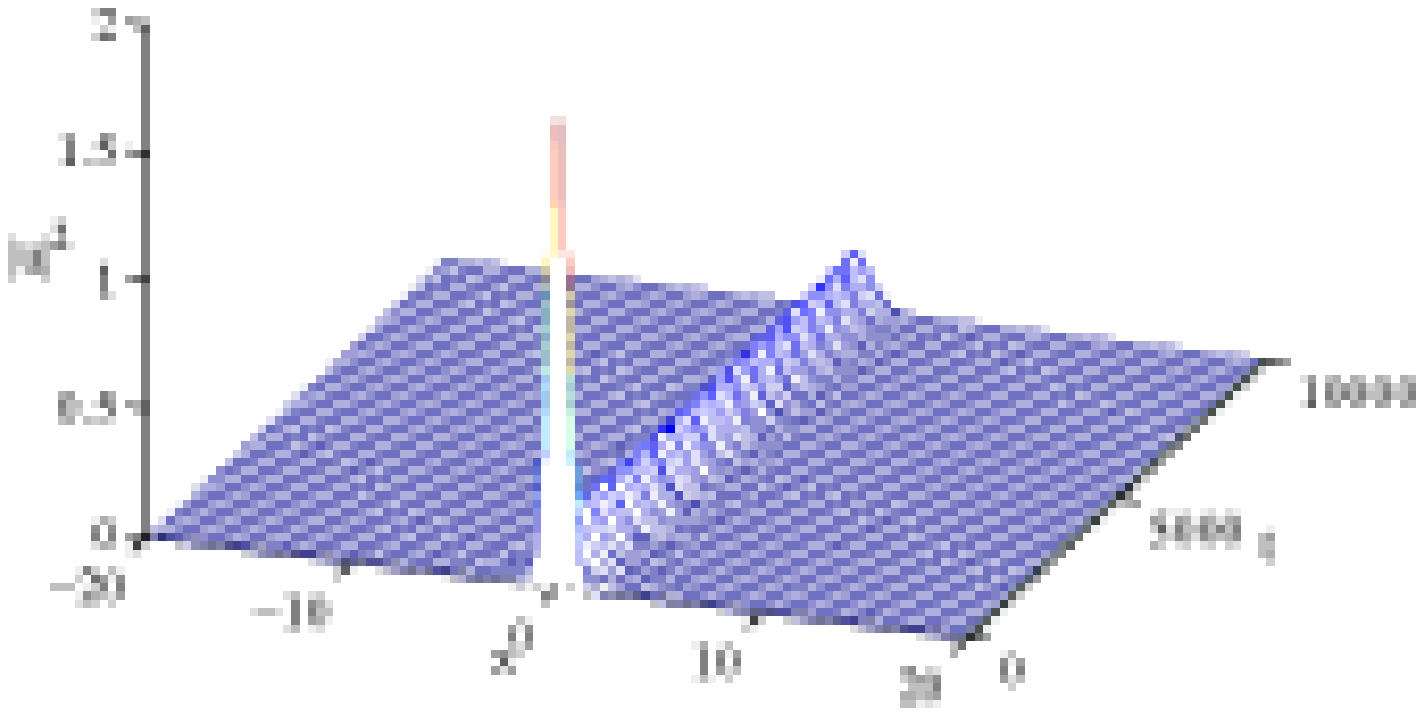}}%
\subfigure[]{\includegraphics[width=2.5in]{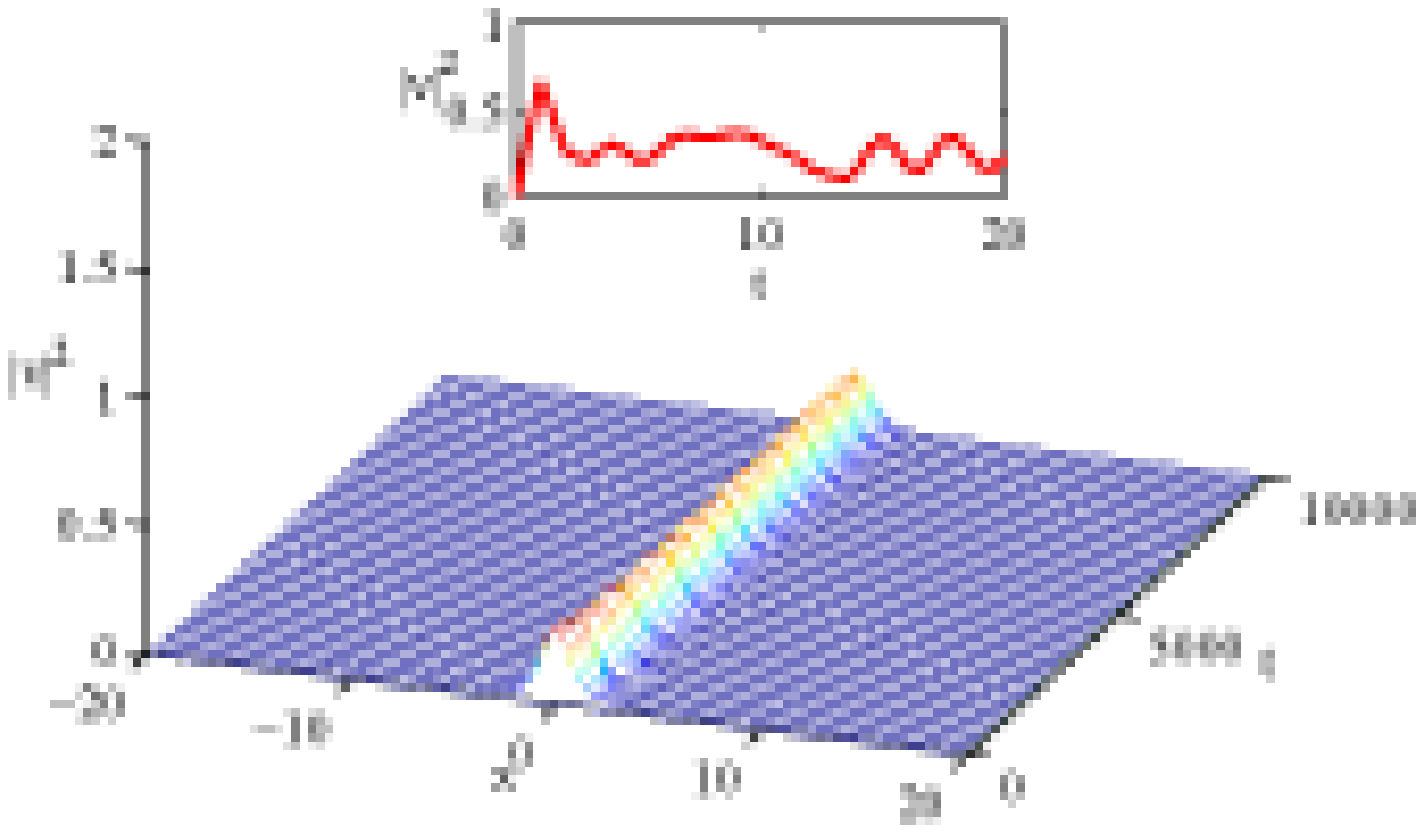}}
\caption{Self-trapping of an input pulse of the forward wave ($u$), at
initial velocity $c=0.2$, into a quiescent ($c=0$) soliton with residual
internal vibrations, which falls into the central bandgap, in the model with
$\protect\mu =0.5$ and $\protect\varepsilon =0$. The inset in (b)
illustrates the initial growth of field $v$, which is absent in the input,
at the soliton's center. The energy of the input pulse is $E=2.28$, of which
$40\%$ is kept by the established soliton.}
\label{fig5}
\end{figure}

Simulations demonstrate that stable \emph{quiescent} single-peak solitons in
the central bandgap can be readily produced by self-trapping of the
one-component moving input pulses, see a typical example in Fig. \ref{fig5}.
In this figure, the velocity of the input pulse is $c=0.2$ (recall $c=1$ is
the largest normalized velocity possible in the model). Faster inputs
generate stable standing solitons with more conspicuous intrinsic
oscillations. It is relevant to mention that the creation of solitons of
``standing light" in fiber BGs is a challenging problem (previously
elaborated theoretical scenarios for that relied on the retardation provided
by a smooth apodization \cite{MakJMO}, and the fusion of colliding solitons
into standing ones \cite{MakPRE}).

\begin{figure}[h]
\centering\subfigure[]{\includegraphics[width=2.5in]{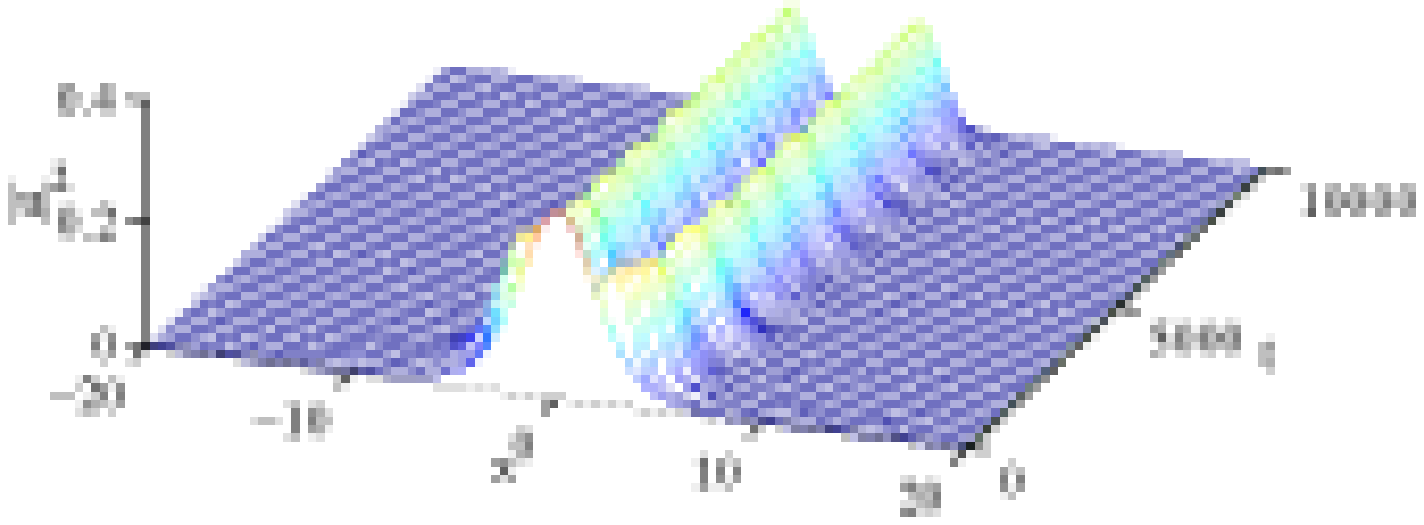}}%
\subfigure[]{\includegraphics[width=2.5in]{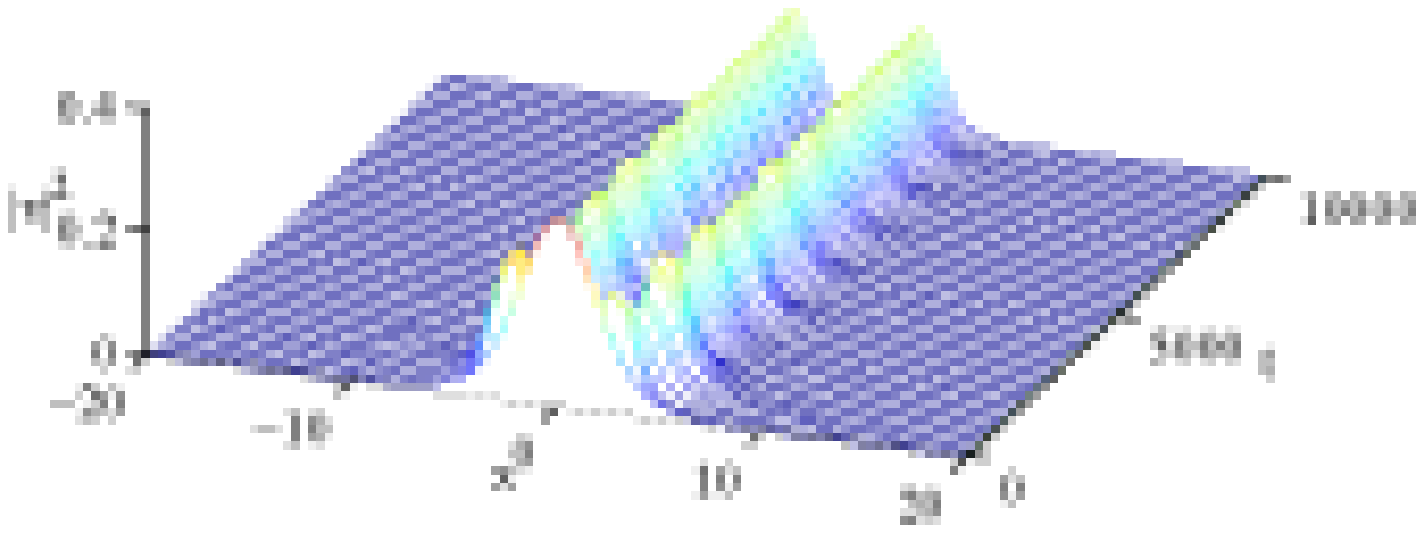}}
\caption{Self-trapping of a two-component input pulse, moving at velocity $%
c=0.2$, into a standing double-peak soliton, in the model with $\protect\mu %
=0.5$ and $\protect\varepsilon =0$. This case is relevant to the
spatial-domain model, see text. The input energy is $E=3.04$, about $60\%$
of which is kept by the emerging double-peak soliton.}
\label{fig6}
\end{figure}

Double-peak GSs belonging to bandgap $-1$ cannot be formed from
single-component inputs, even if the input pulse itself is given a dual-peak
shape. However, they can easily self-trap from moving two-component
single-peak Gaussians, as shown in Fig. \ref{fig6}, in the model with $\mu
>0 $ and $\varepsilon =0$. On the other hand, even small nonzero values of $%
\varepsilon $, if added to this model, make the self-trapping of the
double-peak GSs impossible. This observation stresses, once again, that the
periodic modulation of the chirp (or local refractive index), represented by
$\mu $, generates robust fundamental GSs in gap $-1$, while the reflectivity
modulation, accounted for by $\varepsilon $, strongly destabilizes them. As
mentioned above, the use of the two-component input is possible in terms of
the spatial-BG model.

\subsection{The evolution of unstable solitons}

In the standard BG model, unstable GSs (actually, those with $\omega <0$)
transform themselves into persistent breathers, but they do not demonstrate
re-trapping into stable GSs with a smaller energy. In the present system,
the same is observed as a result of the evolution of unstable solitons in
bandgaps $0$ and $+1$ (not shown here).

\begin{figure}[h]
\centering\subfigure[]{\includegraphics[width=2.5in]{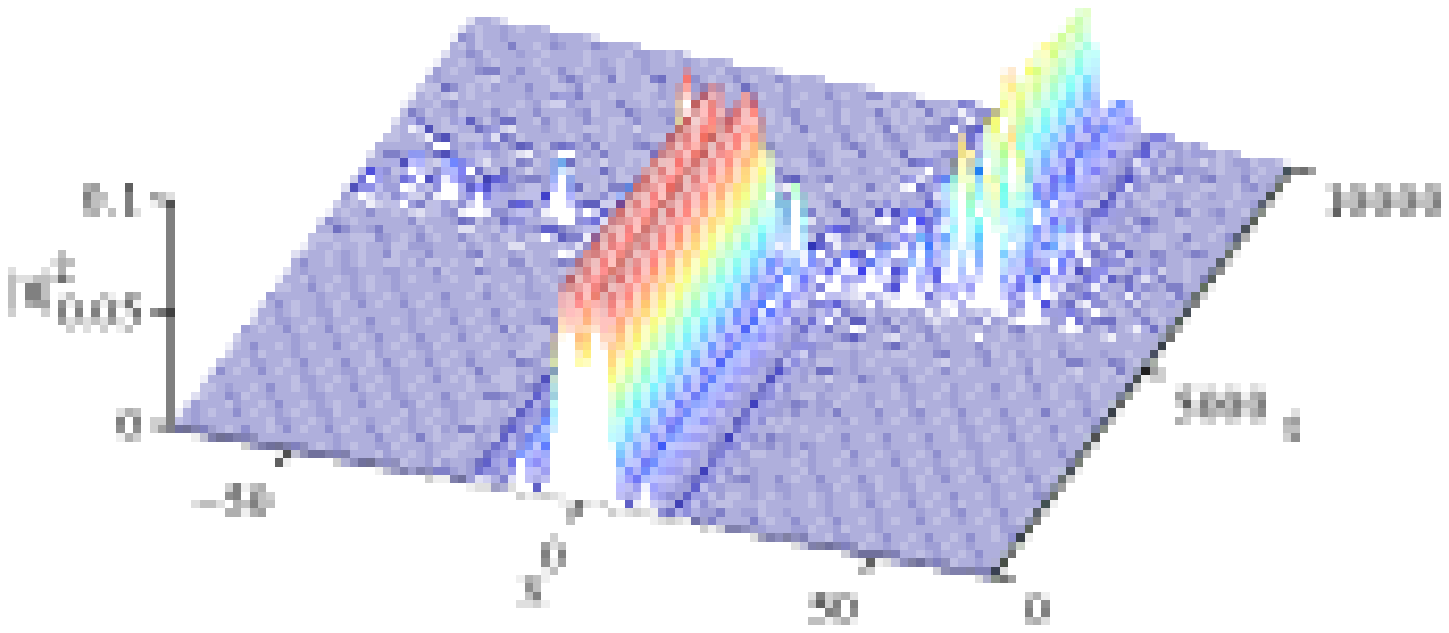}}%
\subfigure[]{\includegraphics[width=2.5in]{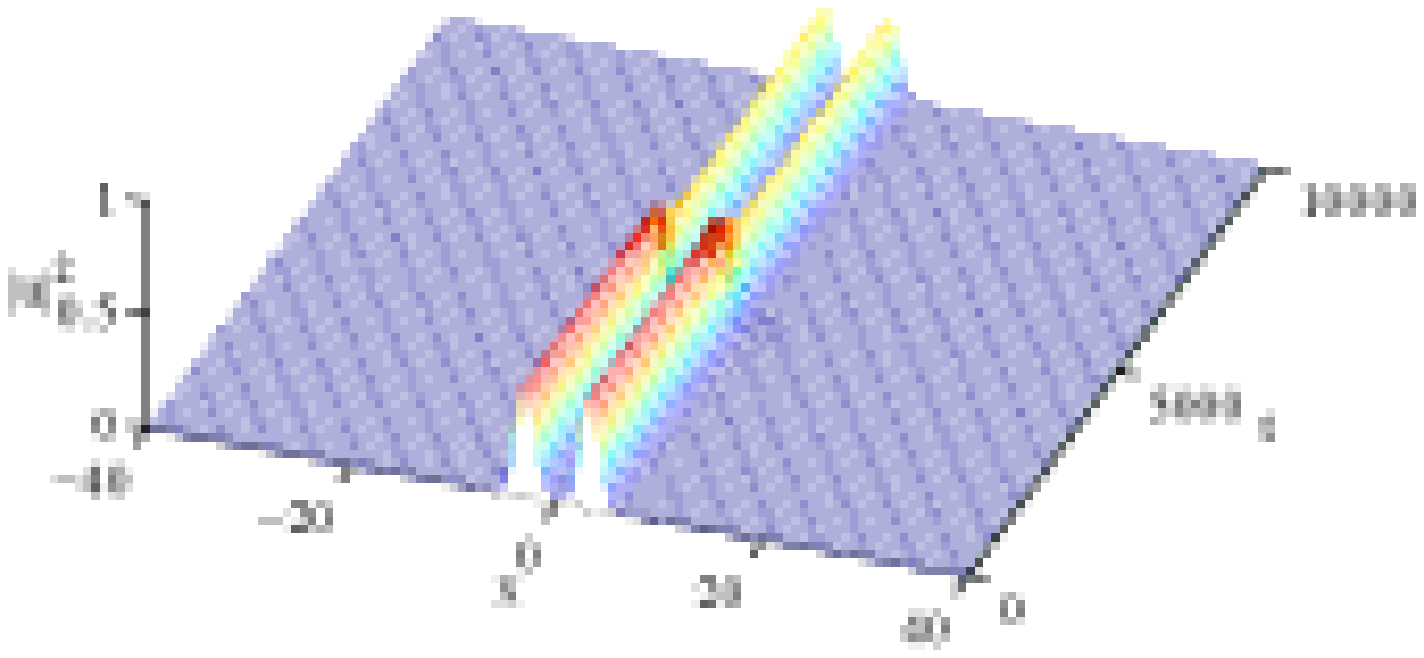}}
\caption{(a) Formation of a breather from an unstable double-peak soliton in
gap $-1$, for $\protect\mu =0.1$, $\protect\varepsilon =0$ and $\protect%
\omega =-1.12$, $E=2.05$. Note the leap of the breather from the original
position. (b) The transformation of an unstable double-peak soliton, with $%
\protect\mu =0.9$, $\protect\varepsilon =0$, $\protect\omega =-1.1$ and $E=5$%
, into a stable gap soliton of the same type, with energy $E=3.3$. In (a)
and (b), only the $u$ component is shown, as the evolution of field $v$ is
quite similar. }
\label{fig7}
\end{figure}

In gap $-1$, unstable solitons with a relatively low energy demonstrate a
more violent instability, which may end up with the formation of a breather
at a position different from that of the original unstable soliton, as shown
in Fig. \ref{fig7}(a). On the other hand, unstable GSs with high energy in
gap $-1$ feature an evolution scenario which does not occur in the standard
model, \textit{viz}., spontaneous rearrangement into another \emph{stable
soliton} belonging to the same bandgap. A typical example of such evolution
is displayed in Fig. \ref{fig7}(b). Unstable double-peak solitons with still
higher energies, which belong to gap $-2$, also self-retrap into stable
two-peak GSs falling into bandgap $-1$.

\subsection{Moving solitons}

As mentioned above, only moving solitons have been observed in experiments
performed in fiber BGs thus far \cite{experiment1,experiment2,slow}. This
fact makes it necessary to study the mobility of stable solitons in the
present model. This was done in the usual way, by applying a \textit{kick}
to stable quiescent solitons, i.e., multiplying them by $\exp \left(
ic_{0}x\right) $.

The double-humped GSs found in gap $-1$ cannot be set in a state of
persistent motion -- they either pass a finite distance and come to a halt,
or get destroyed, if the kink is too strong. On the other hand, stable
single-peak solitons, originally belonging to the central bandgap, can move
at a finite velocity, in the model with $\varepsilon =0$ and small amplitude
of the chirp/refractive index modulation, $\mu \lesssim 0.03$ (moving
solitons practically cannot be created in the model with $\mu =0$ and $%
\varepsilon \neq 0$ \cite{we}) .

\begin{figure}[h]
\centering\includegraphics[width=3in]{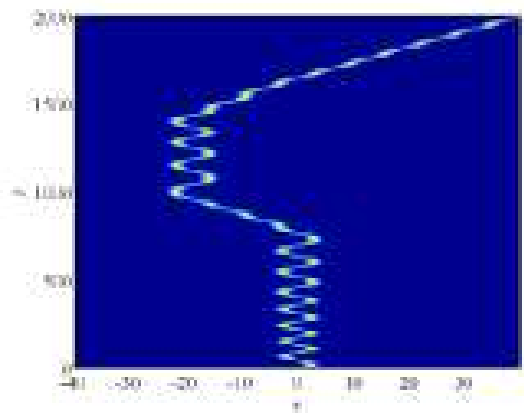}
\caption{Depinning of a soliton with energy $E=3.00$, which belongs to the
central bandgap ($\protect\omega =0.40$) in the model with $\protect\mu %
=0.03 $ and $\protect\varepsilon =0$, by the kick with $c_{0}=0.31$, (this
value only slightly exceeds the depinning threshold). The motion of the
soliton is shown by means of contour plots of $|u|^{2}$.}
\label{fig8}
\end{figure}

To display a generic example of the soliton mobility in the present system,
we notice that, at $\mu =0.03$, the soliton with energy $E=3.00$ remains
pinned if the kick is small, $c_{0}\leq 0.3$. At $c_{0}=0.31$, the kicked
soliton performs several oscillations and then depins itself, starting
progressive motion, as shown in Fig. \ref{fig8}. In this case, the velocity
of the eventual steady motion is found to be $0.12\approx \allowbreak
0.4c_{0}$.

In interval $0.3<c_{0}<0.45$, the same soliton readily sets in persistent
motion, with average velocity $\bar{c}$ which is found to be slightly larger
than $c_{0}/2$ (for example, $\bar{c}=0.19$ for $c_{0}=0.35$). A still
stronger kick sends the soliton in motion for a limited (although long)
interval of time, but then it suddenly gets destroyed by accumulated
disturbances. In the latter case, the velocity observed at the stage of the
quasi-stable motion is much lower than in the truly stable situation, $\bar{c%
}\approx 0.2c_{0}$. On the other hand, if the modulation strength increases
to $\mu =0.05$, kicked GSs do not start to move, but rather demonstrate
oscillations around the pinned state, up to $c_{0}\simeq 0.4$. A stronger
kick destroys them.

\section{Conclusion}

We have reported results of systematic investigation of GSs (gap solitons)
and their moving counterparts in the basic model of periodically modulated
BGs (Bragg gratings), which includes periodic variations of the grating's
chirp (or local refractive index) and reflectivity. In addition to fiber
BGs, the model may also be interpreted in terms of spatial gratings. The
increase of the reflectivity modulation quickly makes all solitons unstable;
on the other hand, the modulation of the chirp supports a new species of
stable BGs in the side bandgap at negative frequencies (gap No. $-1$), and
keeps solitons stable in the central bandgap, No. $0$. The characteristic
feature of the GSs in the side bandgaps is their double-peak shape. The
stable single- and double-peak solitons in gaps $0$ and $-1$, respectively,
demonstrate bistability, existing in overlapping intervals of the energy.
Stable 4-peak bound complexes, formed in bandgap $-1$ by the double-peak
fundamental GSs, were found too.

Quiescent single-peak solitons belonging to the central bandgap readily
self-trap from one-component input pulses, which are launched into the BG at
a finite velocity, while the GSs in gap $-1$ self-trap from the bimodal
input, which is relevant to spatial gratings. On the other hand, unstable
two-peak solitons with a large energy, belonging to bandgaps $-1$ and $-2$,
spontaneously re-trap into stable double-peak GSs (spontaneous rearrangement
of unstable solitons into stable ones does not occur in the standard BG\
model). Moving solitons can be created in the BG with the weak chirp
modulation.

The fabrication of the periodically modulated fiber gratings, considered in
the present model, is quite feasible, and available experimental techniques
should be sufficient for the creation of solitons predicted in this work. In
particular, such experiments may bring closer a solution to the challenging
problem of the creation of solitons made of standing light.

\textbf{Acknowledgements}\smallskip

The work of T.M. is supported, in a part, by a postdoctoral fellowship from
the Pikovsky-Valazzi Foundation, by the Israel Science Foundation through
the Center-of-Excellence grant No. 8006/03, and by the Thailand Research
Fund under grant No. MRG5080171.

\end{document}